\documentclass[12pt,a4paper]{JHEP3} 

\JHEPspecialurl{http://jhep.sissa.it/JOURNAL/JHEP3.tar.gz}

\usepackage{amsmath}
\usepackage{graphicx}
\usepackage{latexsym}
\usepackage{amsfonts}
\usepackage{amssymb}
\usepackage{epsfig}

\newcommand\fverb{\setbox\fverbbox=\hbox\bgroup\verb}
\newcommand\fverbdo{\egroup\medskip\noindent%
            \fbox{\unhbox\fverbbox}\ }
\newcommand\fverbit{\egroup\item[\fbox{\unhbox\fverbbox}]}
\newbox\fverbbox

\newcommand{\be}{\begin{equation}}
\newcommand{\ee}{\end{equation}}
\newcommand{\bea}{\begin{eqnarray}}
\newcommand{\eea}{\end{eqnarray}}

\newcommand{\eref}[1]{(\ref{#1})}

\newcommand{\fref}[1]{Figure~\ref{#1}}

\renewcommand{\t}{\tilde}
\newcommand{\mc}{\mathcal}
\newcommand{\tr}{\rm{tr}}
\newcommand{\Vc}{{\rm Vol}_{{\rm closed}}}
\newcommand{\Vo}{{\rm Vol}_{{\rm open}}}

\newcommand{\drawsquare}[2]{\hbox{%
\rule{#2pt}{#1pt}\hskip-#2pt
\rule{#1pt}{#2pt}\hskip-#1pt
\rule[#1pt]{#1pt}{#2pt}}\rule[#1pt]{#2pt}{#2pt}\hskip-#2pt
\rule{#2pt}{#1pt}}

\newcommand{\fund}{~\raisebox{-.5pt}{\drawsquare{6.5}{0.4}}~}
\newcommand{\antifund}{~\overline{\raisebox{-.5pt}{\drawsquare{6.5}{0.4}}}~}

\newcommand{\symm}{~\raisebox{-.5pt}{\drawsquare{6.5}{0.4}}\hskip-0.4pt%
        \raisebox{-.5pt}{\drawsquare{6.5}{0.4}}~}

\newcommand{\asymm}{~\raisebox{-3.5pt}{\drawsquare{6.5}{0.4}}\hskip-6.9pt%
        \raisebox{3pt}{\drawsquare{6.5}{0.4}}~}

\newcommand{\antiasymm}{~\overline{\raisebox{-3.5pt}{\drawsquare{6.5}{0.4}}\hskip-6.9pt%
        \raisebox{3pt}{\drawsquare{6.5}{0.4}}}~}

\newcommand{\asymmm}{~\overline{\raisebox{-.5pt}{\drawsquare{6.5}{0.4}}\hskip-0.4pt%
        \raisebox{-.5pt}{\drawsquare{6.5}{0.4}}}~}

\title{Gauge theories from D7-branes \\ over vanishing 4-cycles}

\author{Sebasti\'an Franco$^1$, Gonzalo Torroba$^2$ 
\\
\vspace{0.2cm}
~\\
$^1$ Kavli Institute for Theoretical Physics University of California \\
Santa Barbara, CA 93106�4030, USA \\
\vspace{0.2cm}

$^2$ SLAC National Accelerator Laboratory and Physics Department \\
Stanford University, Stanford, CA 94309, USA \\
\vspace{0.2cm}

\email{sfranco@kitp.ucsb.edu, torrobag@slac.stanford.edu}\\
}

\abstract{We study quiver gauge theories on D7-branes wrapped over vanishing holomorphic 4-cycles. We investigate how to incorporate O7-planes and/or flavor D7-branes, which are necessary to cancel anomalies. These theories are chiral, preserve four supercharges and exhibit very rich infrared dynamics. Geometric transitions and duality in the presence of O-planes are analyzed. We study the Higgs branch of these quiver theories, showing the emergence of fuzzy internal dimensions. This branch is related to noncommutative instantons on the divisor wrapped by the seven-branes. Our results have a natural application to the recently introduced F(uzz) limit of F-theory. 
}

\preprint{NSF-KITP-10-130 \\ SLAC-PUB-14283}

\begin{document}

\tableofcontents

\section{Introduction and summary}

Gauge theories with compact seven-branes play a central role in F-theory GUTs (see e.g.~\cite{Donagi:2008ca}) and are basic building-blocks in string compactifications with all moduli stabilized~\cite{Kachru:2003aw}. They may also lead to new types of gauge/gravity dualities~\cite{Heckman:2010pv} and geometric transitions. In this work we investigate quiver gauge theories arising on systems of D-branes at singularities that include D7-branes wrapped on vanishing 4-cycles (``color'' seven-branes). 

A systematic construction and analysis of such theories has not been undertaken yet. This is partly because gauge theories with color D7-branes are in general anomalous. Cancellation of gauge anomalies then requires adding non-compact (``flavor'') D7-branes and/or O7-planes. These gauge theories have rather intricate moduli spaces and nonperturbative dynamics which are not fully understood at present. Gravity solutions capturing these effects are not known. In this work we focus on the gauge theory side, constructing the anomaly-free quivers from cones over toric del Pezzos, with D7 charge. Properties of the associated supergravity solutions will be presented in~\cite{progress1}.

We consider theories on toric singularities that preserve four supercharges; these can be efficiently studied using dimer model techniques \cite{Hanany:2005ve,Franco:2005rj,Franco:2007ii}. First, section \ref{section_quivers_D7s} reviews basic properties of quiver gauge theories with nonzero seven-brane charge. We discuss cancellation of anomalies with O7 planes or D7 flavor branes and present the quiver theory for branes on the Calabi-Yau cone over $\mathbb P^2$. This is the simplest del Pezzo cone, and plays an important role in our work.

Section \ref{section_dP0} presents a detailed analysis of the gauge theory on the Calabi-Yau cone over $dP_0 = \mathbb P^2$, which displays all possible behaviors that arise in more complicated examples. The singular limit is simply the orbifold $\mathbb C^3/\mathbb Z_3$, with a finite $\mathbb P^2$ corresponding to a small resolution. We classify the anomaly-free supersymmetric gauge theories with O7-planes on this background geometry, up to the addition of vector-like matter. We find theories with compact O7-planes (namely, with their internal dimensions wrapped over the vanishing 4-cycle) and others with noncompact orientifolds. Moreover, all these theories are chiral. Based on the results for $dP_0$, we propose a series of rules for identifying and characterizing orientifolds in section \ref{subsection_compact_noncompact_O7s}.

In section \ref{section_dPn} we analyze the orientifolds of the first del Pezzos, $dP_1$, $dP_2$ and $dP_3$. These theories are free of gauge anomalies only after adding flavor D7-branes. Some of these constructions are found to be related by partial resolution. Section \ref{section_nonpert} discusses some nonperturbative effects in quivers with D7-branes and orientifolds. We focus on geometric transitions with non-compact O7-planes and also explain how different orientifold theories are related by Seiberg duality, extending the notion of toric duality \cite{Feng:2001xr,Feng:2001bn} to orientifold theories. 

The theories introduced in this paper have a natural application to the recently introduced F(uzz) theory limit of F-theory \cite{Heckman:2010pv}. Section \ref{sec:Fuzz} is devoted to studying the emergence of fuzzy internal dimensions on the moduli space of quiver theories with D7-branes wrapped over vanishing 4-cycles. This is done both from a four-dimensional and eight-dimensional perspective, connecting noncommutative instantons on a del Pezzo surface to the baryonic branch of the four-dimensional theory. While this work was in preparation, a very nice paper that also explains how fuzzy geometries arise from quivers appeared \cite{Furuuchi:2010gu}. We feel our discussion complements the one in that article, to which we refer the reader for various explicit examples. 

We include an appendix summarizing the rules for orientifolding toric quivers based on dimer models \cite{Franco:2007ii}.

\vskip 1.5mm

Our results constitute a necessary step towards a more complete understanding of gauge theories with color D7-branes at Calabi-Yau singularities. Some of the known phenomena on 5-branes (like the duality cascade~\cite{Klebanov:2000hb} and geometric transitions~\cite{Vafa:2000wi}) may have analogs in theories with 7-branes. In particular, geometric transitions with compact orientifolds are expected to be related to gaugino condensation. It would be interesting to construct supergravity duals of these theories. On the other hand, these seven-brane gauge theories present certain distinctive features: they are chiral and contain 2-index tensor representations under the gauge group. Understanding their nonperturbative dynamics may require new field theory and supergravity techniques which are worth exploring.

\section{Quivers from wrapped D7-branes}\label{section_quivers_D7s}

Placing D-branes at a singularity gives, in the absence of orientifolds, a quiver gauge theory~\cite{Douglas:1996sw} with gauge group
\be
\prod_{i=1}^{n_G}\, U(N_i)
\ee
and matter fields in various (bi)fundamental and adjoint representations. Microscopically, the D-branes at a singularity split into a collection of fractional branes; for four-dimensional theories in IIB, the fractional charges take the general form
$$
F_i=(Q_{7,i},Q_{5,i},Q_{3,i})\;\;,\;\;i=1,\ldots, n_G
$$
where the number of independent fractional branes $n_G$ is equal to the maximum number of gauge groups that a quiver living on D-branes at such singularity can have.

In this work, we focus on the case in which the branes are placed at Calabi-Yau (CY) cones over del Pezzo surfaces. Recall that these are $\mathbb P^2$ blown up at $n$ generic points, $0 \le n \le 8$ (or $\mathbb P^1 \times \mathbb P^1$ blown at $n-1$ generic points). Geometrically, $n_G$ is given by the Euler number of the complex surface being wrapped,
\be
n_G = \chi(dP_n) = n+3\,.
\ee
One of the simplest quivers arises from placing branes on the cone over $\mathbb P^2= dP_0$. Theories including seven-branes in this geometry will be investigated in detail in section \ref{section_dP0}. In this case we have $n_G=3$.

D7 and D5-brane charges can have various components if the geometry has more than one compact 4-cycle or 2-cycle. Fractional brane charges determine the chiral content of the quiver through
\be
N_{ij}={\rm rk} (F_j) \, {\rm deg}(F_i)-{\rm rk} (F_i) \, {\rm deg}(F_j) \, ,
\label{intersection}
\ee
where $N_{ij}$ counts the number of chiral fields in bifundamental representations $(N_i,\bar{N}_j)$, and ${\rm deg}(F_i)=c_1(V_i)\cdot K$, with $K$ the canonical class.

Given a basis of fractional branes $F_i$, the ranks of the gauge groups are related to the total D-brane charges via
\be
(Q_7,Q_5,Q_3)_{\rm tot} = \sum N_i F_i \,.
\ee
Finally, the superpotential of the gauge theory is a polynomial on gauge invariant operators formed from the bifundamental fields (closed, oriented loops in the quiver). Superpotentials for del Pezzo's can be found in
\cite{Feng:2002zw}.

\subsection{Anomaly cancellation: flavoring and orientifolding}

Quivers with wrapped D7-branes (as well as some wrapped D5-branes) suffer from gauge anomalies. An explicit example will be discussed shortly. In order to cancel them, new ingredients have to be added. Two possibilities (which might be introduced simultaneously) are:

\begin{itemize}
\item O7-planes: O7-planes can cancel the charge of the corresponding D7-branes. From a gauge theory point of view, they can turn some of the gauge groups into $Sp/O$ and/or project matter fields into symmetric/antisymmetric representations (or their conjugate).
\item Flavor D7-branes: they introduce matter transforming in the (anti)fundamental representation of certain gauge groups.
\end{itemize}

There is an important difference between the two options. While flavor branes can cancel anomalies introduced by an arbitrary number of D7-branes over vanishing 4-cycles, the allowed number of 7-branes in the case of just O7-planes is always fixed in terms of the orientifold charge.

It is interesting to observe that these anomalies and the resulting constraints also arise in F-theory constructions. Geometrically, the anomaly-free condition comes from tadpole cancellation of twisted fields, localized at the singularity~\cite{Gimon:1996rq}. This is fulfilled in F-theory compactifications on CY four-folds. At weak coupling the tadpole cancellation in the four-fold can be interpreted in terms of orientifolds or anomaly inflow from flavor D7s. This agrees with the previous gauge theory viewpoint.\footnote{We thank D. Morrison for conversations on this point.}

\subsection{Review of the quiver for $dP_0$}

Our analysis will start from the quiver over $dP_0$, so for completeness let us review some basic properties of this theory. The BPS spectrum and properties of this theory were studied in~\cite{Douglas:2000qw}. The CY cone over $\mathbb P^2$ has a degeneration limit where the geometry is $\mathbb C^3/\mathbb Z_3$.

There are three fractional branes, whose charges are \cite{Douglas:2000qw, Franco:2002ae}:
\be
\begin{array}{l}
F_1=(1,0,0) \\
F_2=(-2,H,1/2) \\
F_3=(1,-H,1/2)
\end{array}
\label{fractional_brane_charges_dP0}
\ee
where $H$ is the hyperplane class of $\mathbb P^2$. The total brane charges in this basis,
\begin{equation}
\label{eq:dP0charges}
(N_{D7},\,N_{D5},\,N_{D3}) = \sum_{k=1}^3 N_k\,F_k
\end{equation}
The gauge group ranks are thus given by
\begin{eqnarray}\label{eq:dP0ranks}
N_1&=&N_{D3}+\frac{3}{2}N_{D5}+N_{D7} \nonumber\\
N_2&=&N_{D3}+\frac{1}{2} N_{D5}\nonumber\\
N_3&=&N_{D3}-\frac{1}{2}N_{D5} \,.
\end{eqnarray}

\begin{figure}[ht]
\begin{center}
\includegraphics[width=4.5cm]{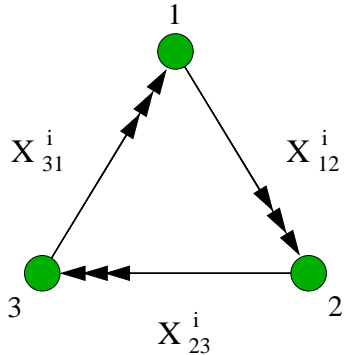}
\caption{Quiver diagram for $dP_0$.}
\label{quiver_dP0}
\end{center}
\end{figure}

The resulting quiver diagram is shown in \fref{quiver_dP0}. The theory has a global $SU(3)$ symmetry under which $X^i_{12}$, $X^i_{23}$ and $X^i_{31}$ ($i=1,2,3$) transform as triplets. The superpotential is
\be
W_{dP_0}=\epsilon_{ijk} X^i_{12} X^j_{23} X^k_{31} \, .
\label{W_dP0}
\ee

\begin{figure}[ht]
\begin{center}
\includegraphics[width=5.5cm]{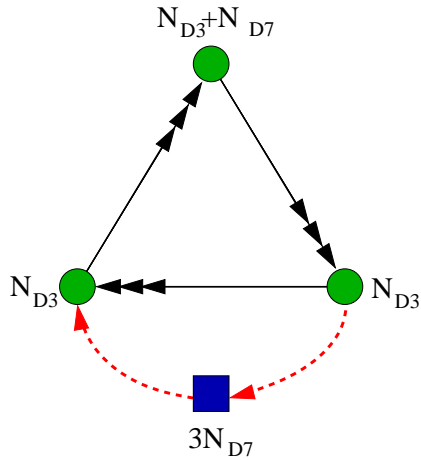}
\caption{One possible anomaly-free quiver diagram for $dP_0$ with $N_{D3}$ D3-branes, $N_{D7}$ D7-branes over the vanishing 4-cycle and $3N_{D7}$ non-compact flavor D7-branes.}
\label{quiver_dP0_flavored}
\end{center}
\end{figure}

For example, consider we want to construct a theory with total D-brane charge $(N_{D7},0,N_{D3})$. Using the charges in \eref{fractional_brane_charges_dP0}, we conclude that the ranks are
$$
N_1=N_{D3}+N_{D7}\;,\;N_2=N_{D3}\;,\;N_3=N_{D3}\,.
$$
This quiver would be anomalous, due to the imbalance between incoming and outgoing arrows at nodes $2$ and $3$. A simple way of cancelling this imbalance is by introducing flavor D7-branes. \fref{quiver_dP0_flavored} shows the resulting anomaly-free quiver. 

There are a variety of holomorphic embeddings of the flavor D7-branes leading to the quiver in \eref{quiver_dP0_flavored} (see e.g. \cite{Ouyang:2003df,Franco:2006es,Franco:2008jc} for some issues regarding flavor D7-branes that are relevant for our discussion). Different embeddings map to different superpotential couplings of the flavors. Also, a discrete choice of Chan-Paton factors controls the pair of nodes in the quiver to which flavors coming from a given stack of flavor D7-branes are connected.

This approach can be straightforwardly implemented for general charges $N_{D7}$, $N_{D5}$, and $N_{D3}$. This class of models is characterized by an arbitrary number of color 7-branes, correlated with the amount of massless matter from (anti)fundamental flavors. The color branes come from D7s wrapped on the vanishing 4-cycle at the base of the cone, while the flavors are noncompact branes extended along the radial direction. F-theory GUT models typically exhibit such flavor branes.

\section{Gauge theory from seven-branes on $dP_0$}
\label{section_dP0}

As discussed in the introduction, cancelling anomalies with orientifolds can give qualitatively different models, without massless fundamental matter. The dynamics of theories with 7-branes is not yet well-understood, making a systematic analysis of gauge theories with D7s and O7s both fruitful and timely. The rest of the paper is devoted to an investigation of these theories and their basic properties.

In this section we will consider the complex cone over $dP_0$, while other toric del Pezzos are analyzed in section \ref{section_dPn}. This example illustrates the entire range of possibilities that will arise in more complicated examples. In addition, identifying the geometric action of the orientifold is particularly simple due to the relation with $\mathbb C^3/\mathbb Z_3$.

\subsection{Fixed point orientifolds}\label{subsec:dP0-pt}

The orientifolds are obtained using the brane tiling methods introduced in \cite{Franco:2007ii} and reviewed in the appendix. They can be divided into orientifolds that lead to fixed points in the brane tiling, and those that produce fixed lines. Let us consider the first class.

\begin{figure}[htb]
\begin{center}     
\includegraphics[width=6cm]{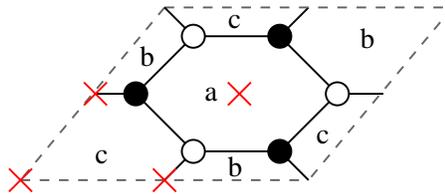}
\caption{Fixed point orientifold for $dP_0$.}
\label{tiling_dP0_points}
\end{center}
\end{figure}

The parity of the orientifold plane is determined by the number of terms in the superpotential. From \eref{W_dP0}, the parity is
$$
\frac{N_W}{2}=1 \;\;\mod (2)\,,
$$
namely we should have an odd number of fixed points with the same sign. The corresponding brane tiling is shown in \fref{tiling_dP0_points}.

Let us denote the face (gauge group) fixed by the orientifold by `a', while `b' and `c' are interchanged by the orientifold action. For this to be a symmetry of the quiver, the corresponding ranks should be equal, $N_b = N_c$. This restricts the allowed brane charges, as we now discuss.

Considering the ranks \eref{eq:dP0ranks}, there are three inequivalent possibilities:\footnote{In what follows, a negative rank denotes anti-branes.}
\begin{enumerate}
\item[{\textit 1)}] $N_a=N_1$, $N_b=N_2$, $N_c=N_3$. This requires $N_{D5}=-\frac{1}{2} N_{D7}$, while $N_{D3}$ is free.
\item[{\textit 2)}] $N_a=N_2$, $N_b=N_3$, $N_c=N_1$. Then $N_{D5}=-N_{D7}$ with $N_{D3}$ arbitrary.
\item[{\textit 3)}] $N_a=N_3$, $N_b=N_1$, $N_c=N_2$, in which case $N_{D5}=0$ with $N_{D3}$ and $N_{D7}$ not fixed at this stage.
\end{enumerate}
Under the orientifold identification, the $X^i_{bc}$ fields are projected to symmetric/antisymmetric representations. We now discuss the 4 inequivalent orientifold parity signs.

\subsubsection*{Compact O7-planes}
\vskip 2mm
\noindent$\bullet\,\,(-\,-\,+\,-)$ 
\vskip 3mm

\noindent Here the notation means that the fixed point in the lower left corner of the unit cell in \fref{tiling_dP0_points} has negative parity and, going clockwise, the second fixed point has parity minus, the third point has parity plus, and the last point has parity minus. In particular, the face `a' has a fixed point with positive parity leading to an SO group.

In the absence of flavor branes, the resulting gauge theory is 
\begin{center}
\begin{tabular}{c|cc}
&$SO(N_a)$&$SU(N_b)$\\
\hline
$X_{ab}^i$&   $\fund$ & $\antifund$ \nonumber\\
$X_{[bc]}^i$&  $1$ & $\asymm$   
\end{tabular}
\end{center}
Anomaly cancellation requires
\begin{equation}
N_a=N_b-4\,.
\end{equation}
The three inequivalent choices above are realized if: {\textit 1)} $N_{D5}=-4, N_{D7}=8$; {\textit 2)} $N_{D5}=4, N_{D7}=-4$; and {\textit 3)} $N_{D5}=0, N_{D7}=-4$. Notice that these theories are chiral.

This suggests the presence of an O7-plane. In order to determine the geometry of the fixed locus we need to find the orientifold action on the mesons of the parent theory, which can be done using the prescription introduced in \cite{Franco:2007ii}.\footnote{It is necessary to point out that the relation between the gauge invariant mesons and normal fluctuations of the branes is in general quite involved. This analysis can be done explicitly here because the geometry is simply $\mathbb C^3/\mathbb Z_3$.} The mesons are given by the invariant combinations
$$
X_{12}^i X_{23}^j X_{31}^k
$$
modulo F-term relations. For $\mathbb C^3/\mathbb Z_3$, we can use coordinates 

\begin{equation}
(z_1,z_2,z_3)\equiv(X_{12}^1 X_{23}^1 X_{31}^1,X_{12}^2 X_{23}^2 X_{31}^2,X_{12}^3 X_{23}^3 X_{31}^3)\,.
\end{equation} 
The involution $\sigma$ on these mesons is
\begin{equation}\label{eq:compact1}
\sigma(z_1,z_2,z_3) = - (z_1, z_2, z_3)\,.
\end{equation}
This corresponds to a compact orientifold, namely one which is compact in the internal dimensions.

After blowing-up the vanishing four-cycle into a $\mathbb P^2$ of finite size, Eq.~(\ref{eq:compact1}) implies that the vanishing locus coincides with $\mathbb P^2$. This unambiguously identifies the orientifold as wrapping the same cycle wrapped by the seven-branes, and the RR tadpole is cancelled locally.

\bigskip
\noindent\underline{Adding flavors}
\medskip

We can also include flavor D7-branes. The matter content becomes

\begin{center}
\begin{tabular}{c|cc}
&$SO(N_a)$&$SU(N_b)$\\
\hline
$X_{ab}^i$&   $\fund$ & $\antifund$ \nonumber\\
$X_{[bc]}^i$&  $1$ & $\asymm$ \\ \hline
$Q_I$ & $1$ & $\fund$  
\end{tabular}
\end{center}
with $I=1,\ldots,F$, with the understanding that $F<0$ corresponds to antifundamentals of $SU(N_b)$. The theory is anomaly-free if
\be
F=3(4+N_a-N_b) \,.
\ee
In particular, we can set $N_a=N_b$ by taking $F=12$. In this case there is no wrapped D7-brane charge and the RR-tadpole of the compact O7-plane is entirely cancelled by the flavor D7-branes. 

Flavor branes can be added to all the examples that follow with similar results. Having illustrated their effect, we focus on the case without flavors unless they are strictly necessary.

\vskip 4mm
\noindent$\bullet\,\,(+\,+\,-\,+)$ 
\vskip 3mm

\noindent A similar theory is obtained by an overall sign change. In this case we have\footnote{Our convention is $Sp(N)\subset SU(N)$ so that, in particular, $Sp(2) \approx SU(2)$.}
\begin{center}
\begin{tabular}{c|cc}
&$Sp(N_a)$&$SU(N_b)$\\
\hline
$X_{ab}^i$& $\fund$ & $\antifund$   \nonumber\\
$X_{\{bc\}}^i$& $1$ & $\symm$
\end{tabular}
\end{center}
where now anomaly cancellation sets
\begin{equation}
N_a = N_b + 4\,.
\end{equation}
For instance, case {\textit 1)} above now requires $N_{D7}=-8$ and $N_{D5}=4$. Similarly, case {\textit 3)} is obtained for $N_{D7}=4$ and $N_{D5}=0$.

The action on the mesons $(z_1,z_2,z_3)$ is as before, corresponding to a compact O7.

\subsubsection*{Non-compact O7-planes}

The other two possible parity assignments give noncompact orientifold-planes, as we now explain.

\vskip 4mm
\noindent$\bullet\,\, (-\,-\,-\,+)$
\vskip 3mm

\noindent Consider this sign choice, as well as two equivalent permutations leaving the third parity sign fixed. The projection gives
\begin{center}
\begin{tabular}{c|cc}
&$Sp(N_a)$&$SU(N_b)$\\
\hline
$X_{ab}^i$& $\fund$ & $\antifund$ \nonumber\\
$X_{[bc]}^i\;(i=1,2)$& $1$ & $\asymm$    \nonumber\\
$X_{\{bc\}}^3$& $1$ &$\symm$    
\end{tabular}
\end{center}
The anomaly cancellation condition
\begin{equation}
3 N_a = 3N_b-4
\end{equation}
cannot be satisfied, since ranks are integer numbers.

The theory is made anomaly-free by adding fundamental flavors for $SU(N_b)$
$$
Q_I\;\;,\;\;I=1,\ldots,4+3 F
$$
and setting
\be
N_a = N_b + F\,.
\ee
For instance, choosing faces as in case {\textit 3)} above gives $N_{D7}=F$ color D7-branes, and $N_{D5}=0$.

The brane system has a noncompact O7-plane, as can be seen from the orientifold action on the $\mathbb C^3/\mathbb Z_3$ coordinates
\be
\sigma(z_1,z_2,z_3) = (z_1,z_2,-z_3)\,.
\ee
The fundamental flavors correspond to non-compact D7-branes required to cancel the RR tadpole of the noncompact orientifold.

\vskip 4mm
\noindent$\bullet\,\,(-\,+\,+\,+)$
\vskip 3mm

\noindent Changing the overall sign, the anomaly free gauge theory becomes
\begin{center}
\begin{tabular}{c|cc}
&$SO(N_a)$&$SU(N_b)$\\
\hline
$X_{ab}^i$& $\fund$ & $\antifund$ \nonumber\\
$X_{\{bc\}}^i\;(i=1,2)$& $1$ & $\symm$    \nonumber\\
$X_{\{bc\}}^3$& $1$ &$\asymm$    \nonumber\\
\hline
$\tilde Q_I$& $1$ &$\antifund$
\end{tabular}
\end{center}
where $\tilde Q_I$, $I=1,\,\ldots,4+3F$ are antifundamental flavors from noncompact D7-branes. They play the same role as in the previous example. The geometric action is

\be
\sigma(z_1,z_2,z_3) = (z_1,-z_2,z_3)\,.
\ee

\subsection{Fixed line orientifolds}\label{subsec:dP0-line}

\begin{figure}[htb]
\begin{center}     
\includegraphics[width=6cm]{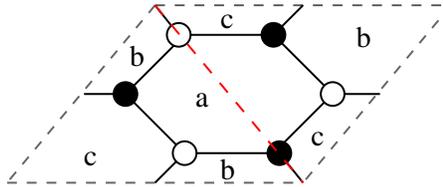}
\caption{Fixed line orientifold for $dP_0$.}
\label{tiling_dP0_line}
\end{center}
\end{figure}

The brane tiling for $dP_0$ also allows for orientifolds associated with fixed lines, as shown in \fref{tiling_dP0_line}. As before, the face fixed by the involution is denoted by `a', while `b' and `c' are interchanged. 

The fixed line can have either positive or negative sign. Consider the case of a positive projection, giving rise to an $SO$ group associated to $N_a$. The fields $X_{bc}$ are projected to a symmetric representation, while $X_{ca}$ and $X_{ab}$ are interchanged and identified. The resulting gauge theory is
\begin{center}
\begin{tabular}{c|cc}
&$SO(N_a)$&$SU(N_b)$\\
\hline
$X_{ab}^i$& $\fund$ & $\antifund$ \nonumber\\
$X_{\{bc\}}^i$& $1$ & $\symm$   
\end{tabular}
\end{center}

This is anomaly-free for
\begin{equation}
N_a= N_b+4\,.
\end{equation}
For instance, case {\textit 2)} has arbitrary $N_{D3}$ and the same number of anti D5-branes and D7s, $-N_{D5}=N_{D7}=4$; case {\textit 3)} gives $N_{D7}=4$ without D5-branes.

The O7-plane in this model illustrates an interesting feature. Using the rules in \cite{Franco:2007ii}, we can determine the orientifold geometric action
\be
\sigma(z_1,z_2,z_3) = (z_3,z_2,z_1)
\ee
and the fixed locus is the noncompact surface $z_1+z_3=0$. Therefore the localized RR-tadpoles from the (compact) color D7-branes cancel against a non-compact O7-plane. This is the counterpart of what we have seen in examples in the previous section, for which the RR-tadpoles from compact O7-planes can be cancelled by non-compact D7-branes. Of course, the orientifold tadpole can also be cancelled just in terms of noncompact D7-branes. For instance, we get an anomaly-free theory with $N_a=N_b$ (so that $N_{D7}=0$) and 12 antifundamentals under the SU subgroup.

\subsection{Identifying and characterizing O7-planes }

\label{subsection_compact_noncompact_O7s}

Based on the previous analysis, let us discuss different approaches for identifying and characterizing O7-planes.
The RR-charge of an O7-plane is $Q_{O7}=\pm 4 Q_{D7}$. The number of D7-branes (compact and non-compact) is constrained accordingly. The presence of $\pm 4$ contributions in the gauge theory comes exclusively from the contribution to anomalies of (anti)symmetric representations of unitary gauge groups. We are then lead to a simple gauge theoretic rule: 
\begin{itemize}
\item A necessary condition for having O7-planes is the presence of some (conjugate) (anti)symmetric representation of a unitary gauge group in the spectrum.
\end{itemize}
From the $dP_0$ example, we see that we can distinguish O7-planes depending on whether their worldvolume is compact or not. A possible way to differentiate both situations is to use the rules in \cite{Franco:2007ii} to identify the geometric action of the orientifold in terms of mesonic operators, as done in the previous sections. While these rules apply to arbitrary toric singularities, the implementation of this approach can be difficult for complicated geometries.

Compact and non-compact O7-planes can also be distinguished using the fractional brane charges for calculating the total D-brane charge of a given configuration. Once again, dimer models provide a general prescription for determining fractional brane charges for arbitrary toric singularities \cite{Hanany:2006nm}.

Partial resolution is another useful tool. From a gauge theory perspective, it corresponds to turning on expectation values for some matter fields, with the resulting higgsing of gauge groups and integration of massive fields.  In some cases, we can use this approach to establish a connection with simple theories, such as orientifolds of $dP_0$, in which the determination of the geometric action of the orientifold is straightforward. Non-compact O7-planes remain non-compact under partial resolution, which thus serves as a way of identifying them.

For $dP_0$ and some of the examples in the next section we see that, in some cases, anomalies cannot be cancelled without the addition of flavors. Then, it is natural to propose that: 
\begin{itemize}
\item If anomalies cannot be cancelled without the addition of fundamental flavors, then the O7-planes are non-compact. 
\end{itemize}
It is important to notice that, as we have seen in explicit examples for $dP_0$, there are cases in which the RR-tadpoles (and hence gauge anomalies) of compact O7-planes can be entirely cancelled by (non-compact) flavor D7-branes. Conversely, there are theories where the RR-tadpoles of non-compact O7-planes can be cancelled by (compact) color D7-branes.

\section{Orientifolds of del Pezzo theories}
\label{section_dPn}

We now present various orientifolds of del Pezzo theories. Many explicit examples have appeared before in the literature (see for example \cite{Franco:2007ii}). The list of models can be taken as reference for future applications. We will use some of them to illustrate different ideas in the coming sections.

\subsection{Orientifolds of $dP_1$}

Let us first consider the cone over $dP_1$. In this case $h^{1,1}=2$ and color D5-branes can be wrapped around the hyperplane class $H$ or the exceptional divisor $E$. The gauge theory has four gauge groups, and the fractional charges and superpotential can be found e.g. in~\cite{Franco:2002ae,Feng:2002zw}.

\begin{figure}[ht]
\begin{center}
\includegraphics[width=6cm]{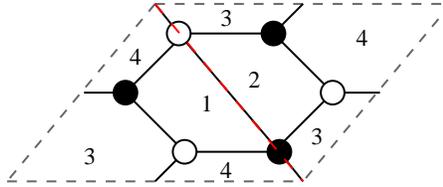}
\caption{Brane tiling for $dP_1$ with the orientifold fixed line.}
\label{tiling_dP1_line}
\end{center}
\end{figure}

$dP_1$ does not admit fixed point orientifolds, but it admits a fixed line orientifold shown in \fref{tiling_dP1_line}.

The matter content is given by

\be
\begin{array}{c|c}
\ \ \ \ \ U_{1/2} \ \ \ \ \ & \ \ \ \ \ U_{3/4} \ \ \ \ \ \\ \hline \hline
 \asymmm/\antiasymm & \\ \hline
 & 3 \times  \asymmm/\antiasymm \\ \hline
\fund & \fund \\ \hline
\fund & \fund \\ \hline
\antifund & \fund
\end{array}
\ee
Even before considering tadpole cancelation, the orientifold identification applied to the fractional branes requires that the brane charges satisfy
\be
N_{D5,E}=0\;\;,\;\;N_{D5,H}+ N_{D7}=0
\ee
where $N_{D5,E}$ and $N_{D5,H}$ denote the number of D5-branes wrapped on the exceptional divisor and hyperplane class, respectively. The ranks of the gauge groups are given by
\be
N_{1/2} = N_{D3}+ \frac{N_{D7}}{2}\;\;,\;\; N_{3/4} = N_{D3}- \frac{N_{D7}}{2}\,.
\ee

The theory is always anomalous because the anomaly-free conditions from the two nodes are
\begin{eqnarray}
N_1-N_3 &=& \pm 4\nonumber\\
N_1-N_3 &=& \mp 4\,.
\end{eqnarray}
The anomaly can be cancelled by adding $\pm 8$ flavors transforming as (anti)fundamentals of $U_{1/2}$.

As we have already mentioned, partial resolution provides an interesting way of connecting orientifolds of different geometries. The orientifolds of $dP_1$ and $dP_0$ given by figures \ref{tiling_dP1_line} and \ref{tiling_dP0_line} are connected by the higgsing associated with removing the edge separating 1 and 2 in \fref{tiling_dP1_line} and relabeling the gauge groups. This entails to give an expectation value to the 2-index tensor of $U_{1/2}$, breaking this to $SO/Sp$ and giving a mass to the fundamental flavors. Using this, the fixed line orientifold of $dP_1$ is identified as noncompact.

Notice that the fixed point orientifolds of $dP_0$ do not appear in the $dP_1$ theory. The wrapped orientifold planes associated to the fixed points provide an obstruction to blowing-up the exceptional $\mathbb P^1 \subset \mathbb P^2$.

\subsection{Orientifolds of $dP_2$}

There are two possible gauge theories for $dP_2$, which are related by Seiberg duality \cite{Feng:2002zw}. We refer to them as Models I and II. While Model I does not admit orientifolding, Model II has a fixed line orientifold, which we analyze below. This is not surprising. In general, some orientifolds might not be present in some of the dual phases. As explained in the appendix, this is because the Seiberg duality transformations that connect the parent phases might not be present due to the orientifold.

\begin{figure}[ht]
\begin{center}
\includegraphics[width=2.5cm]{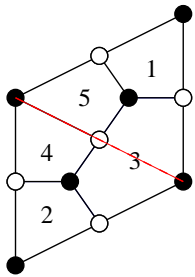}
\caption{Brane tiling for Model II of $dP_2$ with an orientifold fixed line.}
\label{tiling_dP2_II}
\end{center}
\end{figure}

The brane tiling for this orientifold is shown in \fref{tiling_dP2_II}. The spectrum is

\be
\begin{array}{c|c|c}
\ \ \ \ \ U_{1/2} \ \ \ \ \ & \ \ \ \ \ U_{4/5} \ \ \ \ \ & \ SO/Sp_{3} \ \\ \hline \hline
\fund & \antifund &  \\ \hline
\fund & \fund &  \\ \hline
\antifund &  & \fund \\ \hline
\antifund &  & \fund \\ \hline
& \antifund & \fund \\ \hline
& \symm/\asymm &   
\end{array}
\ee
Anomaly cancellation reads
\be
\begin{array}{rcl}
N_3 & = & N_4 \\
N_3 & = & N_4 \pm 4 
\label{anomalies_dP2_II}
\end{array}
\ee	
Once again, it is not possible to satisfy these equations without adding flavors. This is an indication that this is a non-compact O7-plane.

\subsection{Orientifolds of $dP_3$}

There are four dual phases for $dP_3$  \cite{Feng:2002zw}. It is straightforward to see that Models III and IV of $dP_3$ do not admit orientifolding (with either fixed points or fixed lines). 

\subsubsection*{Model I: Orientifold 1}

\begin{figure}[ht]
\begin{center}
\includegraphics[width=2.5cm]{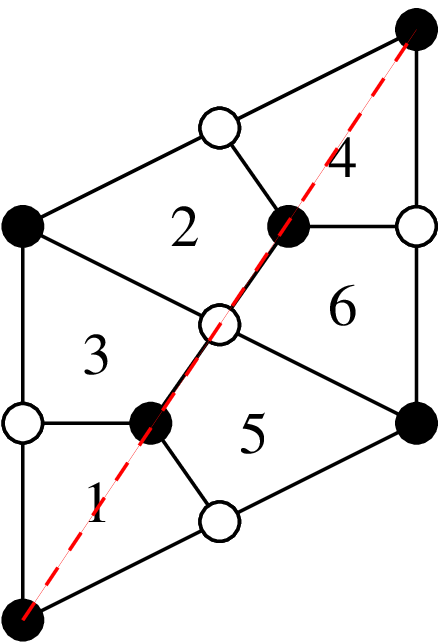}
\caption{Brane tiling for Model I of $dP_3$ with an orientifold fixed line.}
\label{tiling_dP3_I_1}
\end{center}
\end{figure}

This model does not have orientifolds with fixed points. It has two possible fixed lines. We discuss one of them below and the other one in the next subsection. \fref{tiling_dP3_I_1} shows a unit cell in the brane tiling for Model I of $dP_3$ along with one possible orientifold line.

The resulting spectrum is

\be
\begin{array}{c|c|c|c}
 \ SO/Sp_1 \ &  \ SO/Sp_4 \  & \ \ \ \ \ U_{2/6} \ \ \ \ \ & \ \ \ \ \ U_{3/5} \ \ \ \ \ \\ \hline \hline
 \fund &  & \antifund & \\ \hline
 \fund &  & & \antifund \\ \hline
  & \fund & \fund &  \\ \hline
  & \fund & & \fund \\ \hline
 &  & \asymmm/\antiasymm & \\ \hline
 & & & \symm/\asymm \\ \hline
& & \fund & \antifund 
\end{array}
\label{spectrum_dP3_I_1}
\ee
Anomaly cancellation for the unitary gauge groups reads
\be
\begin{array}{rcl}
-N_1-N_2+N_3+N_4 & = & \pm 4 \\
-N_1-N_2+N_3+N_4 & = & \mp 4  
\end{array}
\ee
We conclude that, in the absence of fundamental flavors, there is no solution to the anomaly cancellation equations. This suggests that this configuration contains a non-compact O7-plane.

\subsubsection*{Model I: Orientifold 2}

\label{section_dP3_I_2}

\begin{figure}[ht]
\begin{center}
\includegraphics[width=2.5cm]{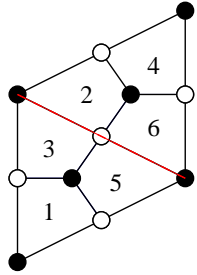}
\caption{Brane tiling for Model I of $dP_3$ with an orientifold fixed line.}
\label{tiling_dP3_I_2}
\end{center}
\end{figure}

The second orientifold of Model I of $dP_3$ corresponds to the fixed line in \fref{tiling_dP3_I_2}.

The matter spectrum is

\be
\begin{array}{c|c|c}
\ \ \ \ \ U_{1/4} \ \ \ \ \ & \ \ \ \ \ U_{2/3} \ \ \ \ \ & \ \ \ \ \ U_{5/6} \ \ \ \ \ \\ \hline \hline
\fund & \fund & \\ \hline
\fund & \antifund & \\ \hline
\antifund & & \fund \\ \hline
\antifund & & \antifund \\ \hline
 & \antifund & \antifund \\ \hline
& \symm/\asymm & \\ \hline
&  & \symm/\asymm \\
\end{array}
\ee
The anomaly cancellation conditions read
\be
\begin{array}{rcl}
N_2 & = & N_5 \\
N_2 & = & N_5 \mp 4 \\
N_2 & = & N_5 \pm 4
\label{anomalies_dP3_I_3}
\end{array}
\ee	
and $N_1$ is arbitrary. Without introducing fundamental flavors, there is no solution to the anomaly cancellation equations. 

\subsubsection*{Model II}

\begin{figure}[ht]
\begin{center}
\includegraphics[width=5cm]{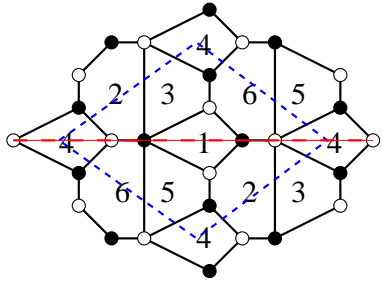}
\caption{Brane tiling for Model II of $dP_3$ with an orientifold fixed line.}
\label{tiling_dP3_II}
\end{center}
\end{figure}

We can also find O7-planes in a dual phase of the $dP_3$ theory, the so called Model II. The brane tiling is shown in \fref{tiling_dP3_II}. We have relabeled gauge groups with respect to \cite{Franco:2005rj}, in order to simplify later comparison (in section \ref{subsection_dual_orientifolds}) with one of the orientifolds of Model I we have just discussed.

The spectrum is

\be
\begin{array}{c|c|c|c}
 \ SO/Sp_1 \ &  \ SO/Sp_4 \  & \ \ \ \ \ U_{2/6} \ \ \ \ \ & \ \ \ \ \ U_{3/5} \ \ \ \ \ \\ \hline \hline
 \fund &  & \fund & \\ \hline
 \fund &  & & \fund \\ \hline \hline
  & \fund & \fund &  \\ \hline
  & \fund & & \fund \\ \hline
 &  & \asymmm/\antiasymm & \\ \hline
& & \fund & \antifund \\ \hline\hline
& & \asymmm/\antiasymm & \\ \hline
& & \antifund & \antifund
\end{array}
\label{spectrum_dP3_II}
\ee
Again, the theory is anomalous without the addition of noncompact D7-branes:
\begin{eqnarray}
N_1 + N_4 &=& 2 N_2 \pm 8\nonumber\\
N_1 +N_4 & =& 2 N_2\,.
\end{eqnarray}

\vskip 2mm

Again, some of these constructions are related by partial resolution. For example, we can go from \fref{tiling_dP3_I_2} to \fref{tiling_dP2_II} by higgsing the 5 and 6 gauge groups (which corresponds to turning on a vev a 2-index tensor representation in the orientifold) and some obvious relabeling. Hence, it is not surprising that the two orientifolds share some general features. In fact, \eref{anomalies_dP2_II} is equal to \eref{anomalies_dP3_I_3}
after dropping the anomaly equation for the $5/6$ node, which becomes $SO/Sp$ (there is a simple sign mismatch that has to do with the node we are preserving in the projection).

\section{Nonperturbative dynamics}
\label{section_nonpert}

The gauge theories constructed in sections \ref{section_dP0} and \ref{section_dPn} have a very rich moduli space of vacua and infrared dynamics. Properties of the Higgs branch of these quivers are investigated in section \ref{sec:Fuzz}, while here we focus on nonperturbative dynamics. We begin the study of geometric transitions and duality in some of the orientifold theories introduced earlier.

Before proceeding, we note that a common feature of the theories we have found is that perturbatively some of the gauge groups are not asymptotically free. This is related to the differences in ranks of the subgroups from nonzero D7-brane charge. These theories should be UV completed in string theory, for instance by embedding them in a compact F-theory construction. It is then important to understand the extent to which infrared effects in the gauge theories are independent of the UV completion. We leave a detailed analysis of this point for the future.

\subsection{Geometric transitions with noncompact orientifolds}

\label{subsection_geometric_transitions}

\begin{figure}[ht]
\begin{center}
\includegraphics[width=13cm]{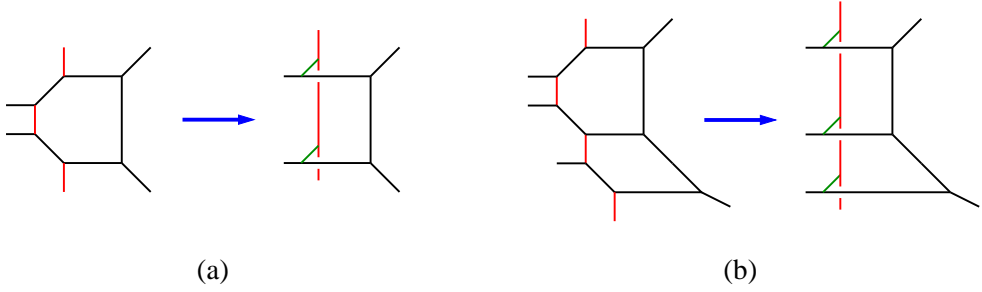}
\caption{Examples of geometric transitions using $(p,q)$ webs: a) one compact 4-cycle disappears, b) two compact 4-cycles disappear.}
\label{geometric_transitions}
\end{center}
\end{figure}

Generalizing what happens for the conifold \cite{Klebanov:2000hb}, geometric transitions map to strong dynamics in the gauge theory \cite{Franco:2005fd}. In a geometric transition one or various compact 2- or 4-cycles disappear. \fref{geometric_transitions} shows some examples of such transitions using $(p,q)$ webs \cite{Aharony:1997ju,Aharony:1997bh}.\footnote{See \cite{Leung:1997tw,Franco:2002ae} for a discussion of the connection between $(p,q)$ webs toric geometry and \cite{Franco:2005fd} for applications in this context.} We now investigate how geometric transitions generalize in the presence of orientifolds. For non-compact O7-planes, we will see that the transition can be regarded as an orientifold projection of the un-orientifolded one. Geometric transitions with compact O7-planes are more subtle, since the compact 4-cycle that supports D7-branes and is the fixed point locus of the orientifold action disappears. This intriguing possibility will be investigated in a future work. Such geometric transitions are potentially interesting for understanding gaugino condensation from a supergravity point of view.

The complex cone over $dP_3$ has a geometric transition to the conifold that has been studied in detail from a gauge theory perspective in \cite{Franco:2005fd}. We now investigate how this transition is modified in the presence of non-compact O7-planes. Let us consider orientifold 2 from section \ref{section_dP3_I_2}. The spectrum reads

\be
\begin{array}{c|c|c|c}
& \ \ \ \ \ U_{1/4} \ \ \ \ \ & \ \ \ \ \ U_{2/3} \ \ \ \ \ & \ \ \ \ \ U_{5/6} \ \ \ \ \ \\ \hline \hline
\ \ X_{1\underline{2}} \ \ & \fund & \fund & \\ \hline
X_{12} & \fund & \antifund & \\ \hline
X_{51}& \antifund & & \fund \\ \hline
X_{\underline{5}1}& \antifund & & \antifund \\ \hline
X_{\underline{2}5}& & \antifund & \antifund \\ \hline
S_2 & & \symm/\asymm & \\ \hline
S_5 & &  & \symm/\asymm \\ \hline
q^i_2 & & \antifund/\fund & \\ \hline
q^i_5 & & & \antifund/\fund \\
\end{array}
\ee
where we have added the (anti)fundamental flavors $q^i_2$ and $q^i_5$ ($i=1,\ldots,4$) that are necessary for anomaly cancellation. The notation for bifundamentals should be understood as follows: any underlined subindex in $X_{ij}$ corresponds to conjugation of the corresponding representation with respect to $(\fund_i,\antifund_j)$. The superpotential reads
\be
W=X_{12}S_2 X_{12} X_{\underline{5}1}S_5 X_{\underline{5}1}+X_{1\underline{2}}X_{\underline{2}5}X_{51}-S_2 X_{\underline{2}5}S_5 X_{\underline{2}5}-X_{1\underline{2}}X_{12}X_{51}X_{\underline{5}1}+W_{flav}
\ee
where $W_{flav}$ denotes couplings between pairs of $q^i_{2,5}$ to other fields in the quiver. There are various possible options for these couplings, depending on the embedding of the flavor D7-branes. Since they do not affect the subsequent discussion, we keep referring to them as $W_{flav}$.

We choose the ranks
\be
N_1=2M\;,\;N_2=N_5=M
\ee
such that $N_{c,1}=N_{f,1}$ and we have a quantum modified moduli space. Physically, this corresponds to $M$ wrapped D5-branes and $M$ D3-branes. Gauge group 1 confines, and the theory is described in terms of mesons 
\be
\mathcal{M}=\left(\begin{array}{ccc} X_{\underline{5}1}X_{1\underline{2}} & \ & X_{\underline{5}1}X_{12} \\
X_{51}X_{1\underline{2}} & \ & X_{51}X_{12} 
\end{array}\right)
\equiv \left(\begin{array}{ccc} M_{\underline{5}\underline{2}} & \ & M_{\underline{5}2} \\ 
M_{5\underline{2}} & \ & M_{52} \end{array}\right)
\ee
and baryons, which we do not write explicitly. Under the remaining gauge symmetry, the mesons transform according to
\be
\begin{array}{c|c|c}
& \ \ \ \ \ U_{2/3} \ \ \ \ \ & \ \ \ \ \ U_{5/6} \ \ \ \ \ \\ \hline \hline
\ \  M_{\underline{5}\underline{2}} \ \ & \fund & \antifund  \\ \hline
M_{\underline{5}2} & \antifund & \antifund  \\ \hline
M_{5\underline{2}} & \fund & \fund \\ \hline
M_{52} & \antifund & \fund  
\end{array}
\ee

The quantum modified moduli space corresponds to
\be
\det \mathcal{M}-\mathcal{B} \tilde{\mathcal{B}}=\Lambda^{4M}
\label{quamtum_ms}
\ee
This modified constraint is implemented using a Lagrange multiplier chiral superfield $X$,
\be
W=S_2 M_{\underline{5}2} S_5 M_{\underline{5}2}+X_{\underline{2}5} M_{5\underline{2}}-S_2 X_{\underline{2}5}S_5 X_{\underline{2}5}-M_{52}M_{\underline{5}\underline{2}}+X (\det \mathcal{M}-\mathcal{B} \tilde{\mathcal{B}}-\Lambda^{4M})+W_{flav}
\ee
We now focus on the mesonic branch, which corresponds to setting $\mathcal{B}=\tilde{\mathcal{B}}=0$ and for simplicity we consider the symmetric situation
\be
\langle \mc M \rangle = M_0\,\mathbf 1_{2M \times 2M}\,.
\ee
This corresponds to non-zero vevs for $M_{\underline{5}\underline{2}}$ and $M_{52}$, which transform as $(\fund,\antifund)$ and $(\antifund,\fund,)$ of $U_{2/3}\times U_{5/6}$, respectively. The Lagrange multiplier is then $X=\Lambda^{4-4M}$.

The gauge group is consequently higgsed to the diagonal subgroup $U_D \subset U_{2/3} \times U_{5/6}$, under which both mesons are neutral. Expanding around the mesonic expectation values, we see that $M_{\underline{5}\underline{2}}$, $M_{52}$, $M_{5\underline{2}}$ and a linear combination of $X_{\underline{2}5}$ and $M_{\underline{5}2}$ are massive. Furthermore, the F-term for $M_{5\underline{2}}$ sets $M_{\underline{5}2} \propto X_{\underline{2}5}$.

The low energy spectrum is
\be
\begin{array}{c|c|c}
& \ \ \ \ \ U(M) \ \ \ \ & \\ \hline \hline
\ \ S_2 \ \ & \symm/\asymm & \ \ S_1/A_1 \\ \hline
S_5 &  \symm/\asymm & \ \ S_2/A_2 \\ \hline
M_{\underline{5}2} & \asymmm+\antiasymm & \bar{S}+\bar{A} \\ \hline
q^i_2,q^i_5 & \antifund/\fund & q^j, j=1,\ldots,8
\end{array}
\ee
where the last column gives a useful nomenclature for the surviving fields. Depending on the charge of the orientifold plane, the superpotential reads
\be
\begin{array}{ccl}
W_+ & = & \bar{A} S_1 \bar{S} S_2 + W_{flav} \\
W_- & = & \bar{A} A_2 \bar{S} A_2 + W_{flav}
\end{array}
\ee
The resulting theory is precisely an orientifold of the conifold, as expected. These orientifolds have been studied in \cite{Park:1999ep} from a IIA Hanany-Witten perspective and in \cite{Franco:2007ii} from a dimer viewpoint.
It is interesting to note that, from a dimer model perspective, this orientifold can be realized both with fixed points and fixed lines. This should not be surprising, since the dimer model orientifolds are basically a practical graphic way of encoding the transformation of the spectrum and interactions under the $\mathbb{Z}_2$ orientifold action.

\begin{figure}[ht]
\begin{center}
\includegraphics[width=8cm]{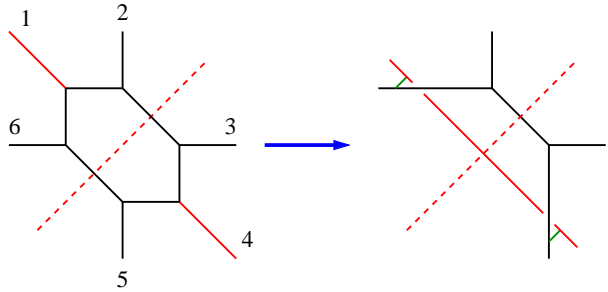}
\caption{$(p,q)$ web representation of the geometric transition from $dP_3$ with an O7-plane to the conifold with an O7-plane.}
\label{geometric_transition_dP3}
\end{center}
\end{figure}

In summary, we have confirmed in an example the expectation that, in the case of non-compact O7-planes, the geometric transition is simply a projection of the unorientifolded one. \fref{geometric_transition_dP3} provides a pictorial representation of this intuition.

\subsection{Dual orientifold theories}
\label{subsection_dual_orientifolds}

In some cases, we have constructed orientifold theories starting from brane tilings of parent theories that are Seiberg dual. Some of the resulting models are also Seiberg dual to each other. This is an extension of toric duality \cite{Feng:2001xr,Feng:2001bn} to theories with orientifolds. In particular, when the duality connecting the parent theories respects the $\mathbb{Z}_2$ symmetry of the orientifold (i.e. when it corresponds to dualizing an arbitrary number of SO/Sp factors and unitary gauge groups with their orientifold images), we obtain orientifolds that are dual to each other. As a result, these theories correspond to exactly the same geometry. A simple example is given by the orientifold of models I and II of $dP_3$ in \eref{spectrum_dP3_I_1} and \eref{spectrum_dP3_II}.

It is straightforward to see that starting from \eref{spectrum_dP3_I_1} and dualizing the $SO/Sp_1$ gauge group, we obtain \eref{spectrum_dP3_II}. Below, we indicate the fields that appear as dual quarks and Seiberg mesons

\be
\begin{array}{cc}
\begin{array}{c|c|c|c}
 \ SO/Sp_1 \ &  \ SO/Sp_4 \  & \ \ \ \ \ U_{2/6} \ \ \ \ \ & \ \ \ \ \ U_{3/5} \ \ \ \ \ \\ \hline \hline
 \fund &  & \fund & \\ \hline
 \fund &  & & \fund \\ \hline \hline
  & \fund & \fund &  \\ \hline
  & \fund & & \fund \\ \hline
 &  & \asymmm/\antiasymm & \\ \hline
& & \fund & \antifund \\ \hline\hline
& & \asymmm/\antiasymm & \\ \hline
& & \antifund & \antifund
\end{array} & 
\begin{array}{l} \\
\left. \begin{array}{c} \\ \\ 
\end{array}\right\} \mbox{dual quarks} \\ \\ \\ \\ \\  
\left. \begin{array}{c} \\ \\
\end{array}\right\} \mbox{mesons} \\ 
\end{array}
\end{array}
\ee
When dualizing, we also generate a meson transforming in the $\asymmm/\antiasymm$ representation of $U_{2/4}$. This field pairs up with one in the conjugate representation that is present in \eref{spectrum_dP3_I_1} and both of them become massive.


\section{Fuzzy geometry on the Higgs branch}
\label{sec:Fuzz}

In the previous sections we have constructed anomaly-free gauge theories with nonzero D7-brane charge including orientifolds and/or flavors. Next we analyze their moduli space of vacua. As usual, there is a Coulomb branch corresponding to meson expectation values that parametrize the orientifolded cone over the del Pezzo surface. Here we will focus instead on the Higgs branch of the theory. While the geometric realization of the Coulomb branch is manifest, the meaning of the Higgs branch in the dual closed string side is much less understood. 

It was argued in~\cite{Maldacena:2009mw} that going into the baryonic branch of the Klebanov-Strassler theory~\cite{Klebanov:2000hb} describes a \textit{fuzzy} $S^2$. When the number of branes becomes large this noncommutative theory gives a good description of the base of the resolved conifold. Our aim here is to explore the analog problem for D7-branes wrapping del Pezzo surfaces, and in particular understand why the Higgs branch is expected to describe a fuzzy version of the internal cycle wrapped by the 7-branes. The essential physics appears already for a flat $\mathbb C^2$ internal manifold, so we will consider this case first in \S\ref{subsec:D3D7}.

In the next step, the local analysis is extended to the cone over the compact 4-cycle. At low energies  we obtain the quiver gauge theories discussed before. The baryonic branch and its relation with the fuzzy geometry will be studied in the 7+1-dimensional theory in \S \ref{subsec:baryonic8d}, and then from the 4d quiver in \S \ref{subsec:baryonic4d}. A very nice general analysis of how to find baryonic branch solutions has recently appeared in~\cite{Furuuchi:2010gu}, so here we will limit ourselves to the simple case of $dP_0$. Our main conclusion will be that the baryonic branch of the quiver associated to 7-branes wrapping a del Pezzo surface is equivalent to noncommutative instantons on this surface. These instantons provide a microscopic interpretation for the appearance of a fuzzy geometry.

\subsection{D3-D7 gauge theory in flat space}\label{subsec:D3D7}

Consider $k$ D3-branes along the directions $(0123)$ and $M$ D7-branes along $(01234567)$, in flat space. The light matter content is 
\begin{center}
\begin{tabular}{c|cc}
&$U(k)$&$U(M)$\\
\hline
&&\\[-12pt]
$Z_1,Z_2$ & {\rm Adj} & $1$ \nonumber\\
$\Phi_1$ & {\rm Adj} & $1$ \nonumber\\
$\Phi_2$ & $1$&{\rm Adj} \nonumber\\
$Q$& $\fund$ & $\antifund$ \nonumber\\
$\tilde Q$& $\antifund$ & $\fund$ 
\end{tabular}
\end{center}
3-3 strings give rise to $U(k)$ gauge fields, two chiral superfields $Z_1$ and $Z_2$ that describe the fluctuations along $(4567)$, and one chiral superfield $\Phi_1$ from motion in the $(89)$ directions. Similarly, 7-7 strings give adjoint $U(M)$ gauge fields and a superfield $\Phi_2$ describing the displacement in $(89)$. Finally, the bifundamentals $(Q, \tilde Q)$ come from 3-7 strings. The full theory is a 7+1-dimensional gauge theory coupled to 3+1 dimensional defects. 

The Higgs branch corresponds to $\Phi_1=\Phi_2=0$. This branch describes how D3s are dissolved into D7s, and it will first be analyzed from the point of view of the effective theory on the D3-branes. The potential is
\be
V \propto {\rm Tr} \left[\left( [Z_1, Z_1^\dag]+[Z_2, Z_2^\dag]+ Q^\dag Q - \t Q \t Q^\dag\right)^2 + \Big|[Z_1,Z_2]+Q \t Q\Big|^2\right]\,.
\ee
The Higgs branch is described as the moduli space of solutions to
\bea\label{eq:ADHM}
\mu_r &\equiv&  [Z_1, Z_1^\dag]+[Z_2, Z_2^\dag]+ Q^\dag Q - \t Q \t Q^\dag =0\nonumber\\
\mu_c & \equiv & [Z_1,Z_2]+Q \t Q=0
\eea
modulo $U(k)$ gauge transformations. 

The moduli space $\mc M_{k,M}$ includes 4 moduli for the positions of the 3-branes inside the 7-brane, plus `size' and `shape' moduli. This is related to the geometry of the internal 7-brane directions --in this case $\mathbb C^2$. However, the field theory is singular at the origin $Z_i= Q = \t Q = 0$, corresponding to the intersection with the Coulomb branch. In order to obtain a well-defined geometric interpretation of the Higgs branch, we need to resolve these singularities.

A simple way of smoothing out the singularity at the origin is to turn on an FI term $\zeta$ for the $U(1)$ part of $U(k)$,
\be\label{eq:FI}
V \propto {\rm Tr} \left[\left( [Z_1, Z_1^\dag]+[Z_2, Z_2^\dag]+ Q^\dag Q - \t Q \t Q^\dag\ - \zeta \,\mathbf 1_k \right)^2 + \Big|[Z_1,Z_2]+Q \t Q\Big|^2\right]\,.
\ee
The moduli space is now
\be\label{eq:curlyM}
\mc M_\zeta = \left(\mu_r^{-1}(\zeta\,\mathbf 1_k) \cap \mu_c^{-1}(0) \right)/U(k)
\ee
and the origin is no longer part of the moduli space. The Higgs branch becomes a smooth manifold. Intuitively, each 3-brane is smeared over a region of size $\zeta$ so the Higgs branch should now realize a ``discretized'' version of the internal manifold. More precisely, we now argue that a fuzzy $\mathbb C^2$ is indeed a subspace of $\mc M_\zeta$.

For this, let us discuss the dynamics from the point of view of the 7+1-dimensional theory, where the D3-branes are $k$ instantons of the $U(M)$ gauge group~\cite{Douglas:1995bn}. The equivalence between both approaches was established in~\cite{Douglas:1996uz}; Eqs.~(\ref{eq:ADHM}) are in fact the ADHM equations. The singularity at the origin corresponds to the small instanton singularity where the D3-branes move off of the D7 worldvolume. The instanton number is given by
\be
k =-\frac{1}{8 \pi^2} \int\, {\rm Tr}(F \wedge F)\,.
\ee
The nonabelian CS action on the D7-brane includes a term
\be
S_{CS} \supset \mu_7 \int C_4 \wedge {\rm Tr} (F \wedge F)
\ee
which, as expected, gives $k$ units of D3-charge. 

The eight-dimensional analog of the FI parameter $\zeta$ corresponds to having instantons on non-commutative $\mathbb R^4$~\cite{Nekrasov:1998ss}. A $B$-field is turned on along $(4567)$, and the relation between the FI term and noncommutativity parameter $\theta$ is
\be
\zeta \sim |\theta|\,.
\ee
The space (\ref{eq:curlyM}) coincides with the solutions to the ADHM equations of noncommutative instantons on $\mathbb R^4$ (see also~\cite{Seiberg:1999vs}). 
In the ADHM equation $\mu_r = \zeta\,\mathbf 1_k$, the FI term is compensated by taking the $Z_i$ to be noncommutative,
\be\label{eq:fuzzyC}
[Z_1, Z^\dag_{\bar 1}] = \theta_{1 \bar 1}\;,\;[Z_2, Z^\dag_{\bar 2}]= \theta_{2 \bar 2}\;\;,\;\;\theta_{1 \bar 1}+\theta_{2 \bar 2}=\zeta
\ee
where $\theta$ has been chosen of type $(1,1)$ in the K\"ahler metric of $\mathbb C^2 \approx \mathbb R^4$. 

Summarizing, the Higgs branch of the D3- D7 system in flat space, in the presence of a nonzero FI term, reproduces a noncommutative version of the $\mathbb C^2$ ``wrapped" by the 7-branes. The FI term comes from a $B$-field in the full 8d theory, and the commutative 4-cycle is recovered in the limit of large $B$. This equivalence between noncommutative instantons in the internal directions and Higgs branch solutions with nonzero FI term in the 4d theory will allow us to understand what happens once the 7-branes wrap a compact four-cycle.

\subsection{The baryonic branch: eight-dimensional perspective}\label{subsec:baryonic8d}

We are now ready to investigate the case of D7-branes wrapping a del Pezzo surface by extending the local the results of \S\ref{subsec:D3D7}. Notice that at low energies the $U(1)$ gauge subgroups of the quiver decouple, becoming global baryonic symmetries. The FI terms in (\ref{eq:FI}) then correspond to nonzero expectation values for certain baryon operators. 

From our previous analysis we thus learn that a baryonic expectation value in the quiver theory maps to a nonzero $B$-field along the internal directions of the 7+1-dimensional theory, and that baryonic branch solutions correspond to noncommutative instantons on the 4-cycle. It will be argued that these instantons are the microscopic degrees of freedom responsible for the appearance of the fuzzy del Pezzo.

The procedure for constructing fuzzy toric geometries starting from (\ref{eq:fuzzyC}) is well-understood -- see for instance~\cite{Balachandran:2005ew} for a review and references. Therefore in this section we study the baryonic branch from the perspective of the 7+1-dimensional theory, explaining the emergence of the fuzzy surface. This will be used in \S \ref{subsec:baryonic4d} to find the corresponding quiver theory with its baryonic branch solutions. 

Although the arguments apply to general toric varieties, for concreteness we focus on the $dP_0$ theory, with anomalies cancelled using noncompact 7-branes. This can be extended to other toric del Pezzos as done recently in~\cite{Furuuchi:2010gu}; while the formulas here already appear in their work, we believe that our approach provides additional insights into the physics of these very rich theories.

\vskip 2mm

We first briefly review the construction of fuzzy $\mathbb P^2$ starting from fuzzy $\mathbb C^3$. Classically, $\mathbb P^2$ corresponds to $z_i \to \lambda z_i$, $i=1,2,3$ and $\lambda \in \mathbb C^*$. Equivalently, the coordinates can be restricted to $S^5$
\be\label{eq:finiteR}
|z_1|^2 + |z_2|^2 + |z_3|^2 = R^2
\ee
and then quotient by $z_i \to e^{i \theta} z_i$. In the quantum version the coordinates are replaced by creation and annihilation operators obeying
\be
[a_i, a^\dag_{\bar j}]= \delta_{i \bar j}\;\;,\;\;[a_i, a_j]=[a^\dag_{\bar i}, a^\dag_{\bar j}]=0\,.
\ee
We further restrict to finite oscillator number,
\be
\sum_i a_i^\dag a_i = n\,,
\ee
which can be thought of as the quantum version of (\ref{eq:finiteR}).

The Fock-space $\mc H_n$ is generated by
\be\label{eq:fock}
|n_1,n_2,n_3 \rangle =\prod_{i=1}^3\,\frac{(a_i^\dag)^{n_i}}{\sqrt{n_i!}} |0\rangle\;\;,\;\;\sum n_i = n\,.
\ee
This has dimension
\be
{\rm dim}\,\mc H_n = \frac{(n+2)!}{2!\,n!}=\frac{1}{2} (n+1)(n+2)\,.
\ee
The $SU(3)$ algebra of $\mathbb P^2$ is realized as usual through the Schwinger construction in terms of the Gell-Mann matrices $\lambda^m$:
\be
L^m = a^\dag_{\bar i} \frac{\lambda^m_{\bar i j}}{2} a_j\;\;\Rightarrow\;\;[L^m, L^n]= i f^{mnr} L^r\,.
\ee
The ``fuzzy coordinates'' are identified with the restriction of the angular momentum operators to $\mc H_n$, with a Casimir operator playing the role of the finite radius (\ref{eq:finiteR}),
\be\label{eq:fuzzy-coords}
X^m \equiv L^m {\big |}_{\mc H_n}\;\;,\;\; \sum (X^m)^2 =  \left(n+ \frac{n^2}{3}\right)\,\mathbf 1{\big |}_{\mc H_n}\,. 
\ee

\vskip 2mm

The eight-dimensional theory consists of $N_{D7}=M$ D7-branes wrapped on $\mathbb P^2$, with nonzero $B$-field proportional to the normalized K\"ahler form $J$,
\be
B = (n+2) J \,.
\ee 
This leads to the periods
\be\label{eq:B}
\int_{\mathbb P^1} B = n+2\;\;,\;\;\int_{\mathbb P^2} B \wedge B = (n+2)^2\,.
\ee
This notation anticipates a relation between the oscillator number $n$ above and the units of magnetic flux, which will be seen shortly. These periods correspond to dissolved D5 and D3 charge respectively. 

The dynamics along the internal cycle is given by noncommutative $SU(M)$ SYM with instanton number
\be
-\frac{1}{8 \pi^2} \int_{\mathbb P^2} {\tr}\,\hat F \wedge \hat F=\frac{1}{2}(n+2)^2\,.
\ee
Here the noncommutative field-strength is given by
$$
\hat F_{i \bar j} \equiv \partial_i \hat A_{\bar j} - \bar \partial_j \hat A_i - [\hat A_i, \hat A_{\bar j}]\,.
$$
The gauge theory is defined on a noncommutative $\mathbb P^2$, given by (\ref{eq:fock})--(\ref{eq:fuzzy-coords}).

\subsection{Constructing the quiver and baryonic branch solutions}\label{subsec:baryonic4d}

These results can be used to determine the 4d quiver theory on $\mathbb P^2$ and the baryonic branch solutions that describe the fuzzy surface. This theory was discussed in \S \ref{section_quivers_D7s}. The first step is to calculate the ranks of the gauge groups in terms of the eight-dimensional induced charges from (\ref{eq:B}). This gives
\be\label{eq:ND3}
N_{D7} = M\;\;,\;\;N_{D5} = (n+2)M\;\;,\;\;N_{D3}=\frac{1}{2} (n+2)^2M\,.
\ee
Using Eq.~(\ref{eq:dP0ranks}) we find
\bea\label{eq:Nbaryonic}
N_1 & =& \frac{1}{2} (n+3)(n+4) M\nonumber\\
N_2 &=& \frac{1}{2} (n+2)(n+3)M\nonumber\\
N_3 &=& \frac{1}{2} (n+1)(n+2) M \,.
\eea
The theory is made anomaly-free by the addition of flavor D7-branes. In particular, we add $3(n+2)M$ antifundamentals under $U(N_1)$, $3(2n+5)M$ fundamentals of $U(N_2)$, and $3(n+3)M$ antifundamentals of the third group $U(N_3)$.

To avoid too many indices in the expressions below, it is convenient to rename the bifundamentals as
\be
X_{12}^i \to A^i\;,\;X_{23}^i \to B^i\;,\;X_{31}^i \to C^i\;\;,\;\;i=1,2,3
\ee
These bifundamentals encode, in particular, the fields $Z_i$ of the flat space analysis. As we have already shown, these become annihilation operators in the 8d theory with noncommutative instantons. So identifying combinations of $(A,B,C)$ with annihilation operators (appropriately restricted to subspaces of finite oscillator number) should give nontrivial solutions in the baryonic branch.

This is done by noticing that~\cite{Furuuchi:2010gu}
\be
N_1 = ({\rm dim}\,\mc H_{n+2})M\;,\;N_2= ({\rm dim}\,\mc H_{n+1})M\;,\;N_3= ({\rm dim}\,\mc H_{n})M
\ee
so that the bifundamental fields can be seen as maps
\bea
A_i &:& \mc H_{n+2} \otimes \mathbb C^M \to \mc H_{n+1}\otimes \mathbb C^M\nonumber\\
B_i&:&\mc H_{n+1}\otimes \mathbb C^M\to \mc H_n \otimes \mathbb C^M\nonumber\\
C_i&:&\mc H_n \otimes \mathbb C^M \to \mc H_{n+2} \otimes \mathbb C^M\,.
\eea
$A_i$ and $B_i$ then indeed correspond to annihilation operators of finite oscillator number:
\be\label{eq:baryon-vev}
A_i = C_A\,a_i \big|_{\mc H_{n+2}} \otimes \mathbf 1_M\;,\;B_i = C_B\,a_i \big|_{\mc H_{n+1}} \otimes \mathbf 1_M\,.
\ee
The F-terms are then satisfied by setting $C_i=0$; this corresponds to the Higgs branch condition $\Phi=0$ of \S \ref{subsec:D3D7}.

More explicitly the matrix elements are of the form
$$
\langle m_1',m_2',m_3' | A_1 | m_1, m_2, m_3 \rangle = C_A \sqrt{m_1}\,\delta_{m_1',m_1-1}\delta_{m_2',m_2} \delta_{m_3',m_3} \otimes \mathbf 1_M\,,
$$
and similarly for the other fields. Using the operator algebra, these solutions satisfy the D-flat conditions where, as explained above, the FI terms are mapped to expectation values that spontaneously break baryon number. For instance, the D-term condition on node one becomes
\be
\sum_{i=1}^3\, A_i^\dag A_i = |C_A|^2(n+2)\,\mathbf 1_{N_1}\,.
\ee
The expectation value for the operator $\mc U_1 \sim \tr (A^\dag A)$ that is the partner of the baryonic current on node one is $\mc U_1 \sim |C_A|^2$.

Therefore, the baryonic branch of the quiver describes a fuzzy version of the four-cycle where the D7-branes are wrapped. Another test comes from relating the spectrum of the quiver to the KK modes of the noncommutative theory. The relation between the 4d and 8d descriptions allows us to understand the reason for this. The baryonic VEVs translate to a $B$-field along the internal directions (through their role as FI terms), which makes the open string theory noncommutative~\cite{Connes:1997cr}. The $B$-field allows the open strings to `see' the internal geometry, even in the limit of a shrinking cycle.

It would be interesting to explore further consequences of the relation between noncommutative instantons on a del Pezzo surface and the baryonic branch of these quivers. Also, while here we have discussed the moduli space for a theory with noncompact flavors, it would be nice to analyze the anomaly-free quivers with orientifolds and no fundamental matter found in section \ref{section_dP0}.

\subsection{Relation to F(uzz) theory}\label{sec:F(uzz)}

In this last section we explain how the recent F(uzz) theory limit of~\cite{Heckman:2010pv} is related to the quiver gauge theories discussed here.

In models with 7-branes wrapped on compact four-cycles, we may consider three limits:
\begin{itemize}
\item A near brane limit $\alpha' \to 0$ keeping the 4-volume fixed, followed by compactification on the 4-cycle. This gives an effective four-dimensional gauge theory obtained by twisting the seven-brane theory. See e.g.~\cite{Donagi:2008ca}.
\item The F(uzz) limit~\cite{Heckman:2010pv}. Here the closed-string volume of the cycle vanishes, but there is a large amount of dissolved D3 charge. The directions along the 4-cycle are noncommutative and an effective four-dimensional description follows by expanding in fuzzy KK modes.
\item The quiver theory obtained by placing D-branes in the complex cone over the shrinking cycle. This is a gauge theory with product gauge groups of different ranks and matter in fundamental and bifundamental representations.
\end{itemize}

In order to understand the relation between these theories, let us briefly review the F(uzz) proposal. The starting point is the DBI action for the D7-branes wrapped over a 4-cycle $\Sigma$,
\begin{equation}
S_{DBI} = - T_7 \int d^8 \xi\,\sqrt{-{\rm det}(g+ \mc B)}\,,
\end{equation}
with $\mc B = dA + B_2$. The gauge coupling of the effective 4d theory obtained by compactifying the D7-branes on $\Sigma$ is proportional to the volume of the ``open'' string metric $g+B$,
\begin{equation}
\frac{1}{g_4^2}= \frac{\Vo(\Sigma)}{(2 \pi)^3 g_s}\;,\; \Vo(\Sigma) \equiv \Vc(\Sigma) +\frac{1}{2} \int_\Sigma \mc B \wedge \mc B\,.
\end{equation}
The exact decoupling limit $\Vc(\Sigma)=0$ can be taken if a nonzero $B$-field is turned on along the internal cycle, giving
\begin{equation}\label{eq:g4}
\frac{8\pi}{g_4^2}= \frac{1}{(2\pi)^2 g_s}\,\int_\Sigma \mc  B \wedge \mc B=\frac{N}{g_s}\,.
\end{equation}
Here $N \in \mathbb Z$ measures the quantized units of $B$-field.

The proposal of~\cite{Heckman:2010pv} is that this decoupling limit gives rise to a noncommutative theory along the four-cycle, due to the presence of the $B$-field~\cite{Connes:1997cr,Seiberg:1999vs}. This was not derived from a microscopic theory, although it was suggested that it could arise from the Higgs branch of an $SU(N \times M)$ gauge theory, where $M$ is the number of wrapped D7-branes, and $N$ is the dissolved D3-charge (\ref{eq:g4}) per D7-brane.

Our results in \S \ref{subsec:baryonic4d} provide an explicit check for this proposal, showing how the baryonic branch of the del Pezzo quiver reproduces the F(uzz) theory as a low energy limit. In this case, the number of flux units $N$ corresponds to $n^2$ in (\ref{eq:B}) and (\ref{eq:ND3}), in the large $N$ limit. There is also nonzero D5-brane charge $\sqrt{N} M$.

The bifundamental fields $(A,B)$ acquire expectation values (\ref{eq:baryon-vev}) that describe how the D3s and D5s are dissolved into the D7-branes. They reproduce the fuzzy $\mathbb P^2$. The unbroken gauge group $SU(M)$ on the M D7-branes is embedded in the quiver product group with multiplicity $N$,
\be
SU(M)_D \subset SU(M)_1^N \times SU(M)_2^N \times SU(M)_3^N \subset SU(N_1) \times SU(N_2) \times SU(N_3)\,.
\ee
Recall that in the large $N$ limit, $N_i \approx N \times M$ for $i=1,2,3$. The gauge couplings in the original quiver are $1/g^2 \sim 1/g_s$, so the gauge coupling in the diagonal $SU(M)$ becomes $1/g_D^2 \sim N/g_s$. This reproduces (\ref{eq:g4}).

Since the noncommutativity parameter is set by $\mc B^{-1} \sim N^{-1/2}$, at large $N$ we recover the commutative geometry of the internal four-cycle. In particular, the fuzzy KK modes are expected to reproduce the KK modes of the twisted seven-brane theory compactified on a commutative four-cycle of fixed closed string volume. In this sense, the large $N$ limit of the quiver can be interpreted as a deconstruction of the eight-dimensional theory.\footnote{Similar ideas have been explored in theories where the noncommutativity arises from adjoint fields in five-branes. See e.g.~\cite{Andrews:2006aw}.}


\acknowledgments
We thank
M.~Douglas,
B.~Heidenreich,
S.~Kachru,
L.~McAllister,
D.~Morrison,
S.~Sch\"afer-Nameki,
D.~Simic,
A.~Uranga and
H.~Verlinde for useful discussions and comments on the draft. 
S. F. is supported by the National Science Foundation under Grant No. PHY05-51164. G. T. is supported by the US DOE under contract number DE-AC02-76SF00515 at SLAC. G. T. would like to thank the KITP for hospitality, where part of this work was done.

\appendix

\section{Orientifolding brane tilings}

Here we briefly review the results of \cite{Franco:2007ii}. Brane tilings provide a practical tool for studying gauge theories on D-branes on orientifolds of toric singularities. They efficiently encode the identification of gauge groups and matter fields (whose corresponding arrows in the quiver have an orientation flip associated with the worldsheet orientation reversal $\Omega$) that follows from the orientifold projection.

\begin{figure}[ht]
\begin{center}
\includegraphics[width=10cm]{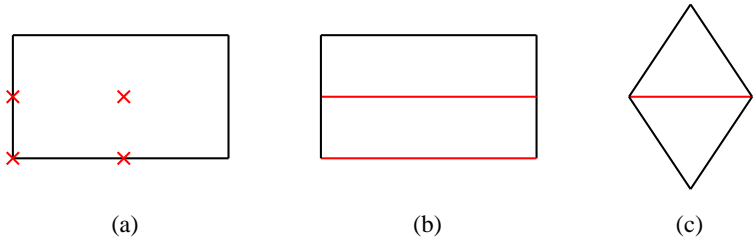}
\caption{Possible orientifolds of brane tilings. a) Orientifold with fixed points. b,c) Orientifold with two or one fixed lines. The $\mathbb{Z}_2$ symmetry of the cases with fixed lines exists only for specific complex structures of $\mathbb{T}^2$.}
\label{tiling_orientifold_possibilities}
\end{center}
\end{figure}

The orientifold action maps to a $\mathbb{Z}_2$ involution of the $\mathbb{T}^2$ in which the tiling is embedded. The involution can correspond to the inversion of two or one of the directions of the $\mathbb{T}^2$. In the first case, the orientifold action has four fixed points. In the second case, we can have one or two fixed lines. All these possibilities are illustrated in \fref{tiling_orientifold_possibilities}.

Each fixed point or line carry an independent sign or charge. In the case of orientifolds with fixed points, the product of the four signs is constrained by supersymmetry to be $(-1)^{N_W/2}$, with $N_W$ the number of superpotential terms in the parent theory. The charges of fixed lines are unconstrained.

Throughout the paper, we refer to the unorientifolded theory as the parent theory. The orientifold theory is obtained by projecting gauge groups and matter fields according to the following rules:

\bigskip

\begin{itemize}
\item {\bf Gauge groups:} the gauge group factor associated to faces that are mapped to themselves is projected down to $SO$ or $Sp$ for positive or negative charge of the fixed point/line, respectively. Faces that are paired up by the orientifold action are identified, giving rise to a single unitary gauge factor.
\item {\bf Matter:} edges mapped to themselves correspond to chiral multiplets in the two-index symmetric or antisymmetric tensor representations for positive or negative charge of the fixed point/line, respectively. Edges paired up by the orientifold action are
identified, giving rise to a single bifundamental field.
\end{itemize}

\bigskip

The superpotential is given by that of the parent theory, written in terms of the projected fields. A more detailed statement of these rules is given in \cite{Franco:2007ii}. Brane tilings are very useful for determining the geometric action of the orientifold, which is encoded in the transformation properties of gauge invariant mesonic operators \cite{Franco:2007ii}. 

The graphic nature of tilings result in rules that simplify the classification of orientifolds. For example, consider orientifolds with fixed points: since the orientifold action must identify nodes of different colors, a necessary condition for an $n$-sided face to contain a fixed point is that $n=2 \, \mbox{mod}(4)$.

\subsection*{Orientifolding Seiberg dual parents}

In general, more than one gauge theory (correspondingly more than one brane tiling) can be associated to a given unorientifolded geometry. These theories, to which we refer as dual phases, are connected by Seiberg duality transformations. Any of these dual theories can be taken as the parent to be orientifolded. The rules explained in the previous section might lead to orientifolds that are present in some phases but absent from others. The reason is simple, the Seiberg dualities that take from one phase to the other might not be symmetric under the orientifold action. 

The simplest example corresponds to two dual phases I and II of the parent theory that are connected by a Seiberg duality on a single gauge group, which we denote $i$. Imagine there is an orientifold of phase I in which gauge group $i$ is not invariant but is mapped to another one $j$. Under these circumstances, we conclude that this orientifold cannot be seen by starting from phase II, since it would require further dualizations. This issue is also discussed in \S\ref{subsection_dual_orientifolds}.


\end{document}